\documentclass[aps,prb,twocolumn,showpacs,preprintnumbers,amsmath,amssymb,superscriptaddress]{revtex4}

\usepackage{graphicx}
\usepackage{dcolumn}
\usepackage{bm}
\usepackage{color}

\newcommand{\e}{\varepsilon}
\newcommand{\s}{\sigma}
\newcommand{\expect}[1]{\langle #1 \rangle}

\begin{document}

\title{Kondo effect in a quantum dot coupled to ferromagnetic leads
and side-coupled to a nonmagnetic reservoir}

\author{I. Weymann}
\email{weymann@amu.edu.pl} \affiliation{Department of Physics,
Adam Mickiewicz University, 61-614 Pozna\'n, Poland}
\affiliation{Physics Department, Arnold Sommerfeld
Center for Theoretical Physics and Center for NanoScience, \\
Ludwig-Maximilians-Universit\"at, Theresienstrasse 37, 80333
Munich, Germany}

\author{J. Barna\'s}
\affiliation{Department of Physics, Adam Mickiewicz University,
61-614 Pozna\'n, Poland} \affiliation{Institute of Molecular
Physics, Polish Academy of Sciences, 60-179 Pozna\'n, Poland}

\date{\today}

\begin{abstract}
Equilibrium transport properties of a single-level quantum dot
tunnel-coupled to ferromagnetic leads and exchange-coupled to a
side nonmagnetic reservoir are analyzed theoretically in the Kondo
regime. The equilibrium spectral functions and conductance through
the dot are calculated using the numerical renormalization group
(NRG) method. It is shown that in the antiparallel magnetic
configuration, the system undergoes a quantum phase transition
with increasing exchange coupling $J$, where the conductance drops
from its maximum value to zero. In the parallel configuration, on
the other hand, the conductance is generally suppressed due to an
effective spin splitting of the dot level caused by the presence
of ferromagnetic leads, irrespective of the strength of exchange
constant. However, for $J$ ranging from $J=0$ up to the
corresponding critical value, the Kondo effect and quantum
critical behavior can be restored by applying properly tuned
compensating magnetic field.
\end{abstract}


\maketitle

\section{Introduction}

Transport through a model single-level quantum dot captures many
interesting and important features of transport phenomena in real
quantum dots. One of such phenomena, which has been of great
interest in the last decade, is the Kondo
effect.~\cite{goldhaber-gordon_98,cronenwett_98,hewson_book} When
the dot is occupied by a single electron, virtual transitions
between the dot and electron reservoirs (external leads) cause
spin fluctuations in the dot. As a result, the dot's spin becomes
screened by electrons of the reservoirs, which results in the
formation of a non-local spin singlet ground state of the system.
Furthermore, a resonance in the density of states appears at the
Fermi level, which gives rise to enhanced transmission through the
dot. In experiments, this leads to the well-known zero-bias
anomaly, i.e. a peak at zero bias in the differential
conductance.~\cite{goldhaber-gordon_98,cronenwett_98}

When the reservoirs are ferromagnetic, the effective exchange
field generated by the electrodes may suppress the Kondo
anomaly.~\cite{LopezPRL03,martinekPRL03_2,martinekPRL03,ChoiPRL04,
MatsubayashiPRB07,SimonPRB07} More specifically, when the dot
described by an asymmetric Anderson model is symmetrically coupled
to ferromagnetic leads, then the Kondo effect becomes suppressed
in the parallel configuration, while in the antiparallel
configuration the Kondo anomaly survives. However, the Kondo
effect in the parallel configuration can be restored, when an
external magnetic field, which compensates the exchange field
created by the ferromagnetic leads, is
applied.~\cite{martinek_PRB05,sindel_PRB07} This behavior was
confirmed in a couple of recent
experiments.~\cite{pasupathy_04,heersche_PRL06,hamayaAPL07,
hamaya_PRB08,hauptmann_NatPhys08,parkinNL08}

The situation becomes more complex and physically richer when the
dot is exchange-coupled to an additional
reservoir.~\cite{OregPRL03} Such a model captures the essential
physics of the so-called two-channel Kondo
effect.~\cite{Nozieres_JP80,Zawadowski_PRL80,AffleckPRB93,RalphPRL94,
HettlerPRL94,AndreiPRL95,LebanonPRB03,FlorensPRL04} In the
two-channel Kondo problem, two separate electron reservoirs
(channels) compete with each other to screen the impurity's spin.
If the coupling to one of them is larger than to the other one, a
usual single-channel Kondo state (spin singlet) is formed between
the dot and more strongly coupled reservoir. This results in two
competing Kondo ground states of the system, depending on the
ratio of coupling strengths to the first and second conduction
channels. Interestingly, these two Kondo states are separated by a
quantum critical point, where both couplings are equal and an
exotic two-channel Kondo state is formed, which cannot be
described within the Landau Fermi-liquid theory. Very recently,
the two-channel Kondo effect has been explored experimentally in
quantum dots.~\cite{PotokNature07} The experimental setup
consisted of a small quantum dot coupled to external leads and to
a large Coulomb-blockaded island. While the electrons could tunnel
between the dot and the leads, only virtual tunneling processes
between the dot and the island were allowed, resulting in an
exchange coupling. By tuning the exchange coupling, it was
possible to study the quantum phase transition between the two
ground states of the system and analyze transport behavior in the
non-Fermi liquid regime.~\cite{PotokNature07} Theoretically, such
a two-channel setup can be modelled for example by a quantum dot
which is tunnel-coupled to external leads and exchange-coupled to
another electron
reservoir.~\cite{PustilnikPRB04,tothPRB07,LiuJPCM08}

As discussed above, both the Kondo effect in a quantum dot coupled
to ferromagnetic leads and the two-channel Kondo phenomenon were
already extensively studied. However, the interplay of leads'
ferromagnetism and two-channel Kondo effect remains to a large
extent unexplored. Therefore, in this paper we address the
two-channel Kondo problem in the presence of ferromagnetism. In
particular, we consider an Anderson quantum dot coupled to
ferromagnetic leads and exchange-coupled to a nonmagnetic electron
reservoir. Using the numerical renormalization group (NRG) method,
we analyze the interplay between the effects due to ferromagnetism
of the leads and exchange coupling to the additional nonmagnetic
reservoir. Depending on the strength of the tunnel coupling $t$
and exchange coupling $J$, the dot's spin can be screened either
by electrons in the ferromagnetic leads or by electrons in the
nonmagnetic reservoir. By analyzing the equilibrium spectral
functions and the conductance through the dot, we show that in the
antiparallel magnetic configuration, the system undergoes a
quantum phase transition with increasing exchange coupling $J$,
where the conductance drops from the maximum value to zero. For a
certain critical value of $J$, $J_{\rm c}^{\rm AP}$, both electron
channels try to screen the dot's spin and the conductance
approaches a half of the quantum conductance. In the parallel
configuration, on the other hand, the conductance is generally
suppressed, irrespective of the exchange constant $J$, due to
effective spin splitting of the dot level caused by the exchange
field coming from ferromagnetic
leads.~\cite{martinekPRL03,ChoiPRL04} We show that the Kondo
effect can be restored by applying a properly tuned external
magnetic field $B$ for $J$ below the corresponding critical point,
$J<\tilde{J}_{\rm c}^{\rm P}$. Furthermore, the quantum critical
regime can also be recovered, which however requires a fine-tuning
in the parameter space of $J$ and $B$.

The paper is organized as follows. In section II we present the
model as well as briefly describe the NRG method together with
some details of calculations. In turn, in section III we present
numerical results for symmetric and asymmetric Anderson models in
both parallel and antiparallel magnetic configurations of the
system. Finally, we conclude in section IV.

\section{Theoretical description}

\subsection{Model}

\begin{figure}[t!]
  \includegraphics[width=0.7\columnwidth]{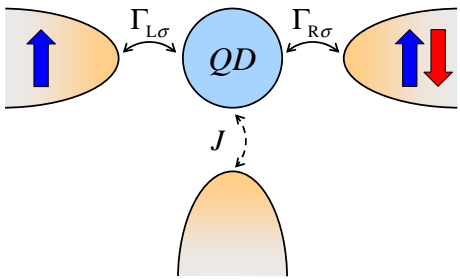}
  \caption{\label{Fig:1}
  (color online) The schematic of a quantum dot (QD) tunnel-coupled
  to external ferromagnetic leads and exchange-coupled to
  a nonmagnetic electron reservoir. The spin-dependent coupling
  to the left (right) lead is described by $\Gamma_{\rm L\s}$
  ($\Gamma_{\rm R\s}$), while $J$ denotes the exchange coupling constant.
  The magnetizations of the leads can form either parallel or antiparallel
  magnetic configuration, as indicated.}
\end{figure}

The considered system consists of a single-level quantum dot
tunnel-coupled to left and right ferromagnetic leads and
exchange-coupled to a nonmagnetic reservoir, see Fig.~\ref{Fig:1}.
It is assumed that the external leads are made of the same
ferromagnetic material and their magnetizations are collinear, so
that the system can be either in the parallel or antiparallel
magnetic configuration. The total Hamiltonian is given by
\begin{equation}\label{Eq:H}
H = H_{\rm FM} + H_{\rm NM} + H_{\rm QD} + H_{\rm tun} + H_{\rm
exch} .
\end{equation}
Here, $H_{\rm FM}$ describes the ferromagnetic leads, $H_{\rm FM}
= \sum_{rk\s} \e_{rk\s} c^\dag_{rk\s} c_{rk\s}$, where
$c^\dag_{rk\s}$ is the electron creation operator with wave number
$k$, spin $\s$ in the left ($r={\rm L}$) or right ($r={\rm R}$)
lead, and $\e_{rk\s}$ is the corresponding energy. The second
part, $H_{\rm NM}$, corresponds to a nonmagnetic electron
reservoir and is given by, $H_{\rm NM} = \sum_{k\s} \e_{k}
a^\dag_{k\s} a_{k\s}$, with $a^\dag_{k\s}$ being the respective
creation operator and $\e_{k}$ is the single-particle energy. The
quantum dot is described by the Anderson Hamiltonian,
\begin{equation}
H_{\rm QD} = \sum_\s \e_{\rm d} d^\dag_\s d_\s + U d^\dag_\uparrow
d_\uparrow d^\dag_\downarrow d_\downarrow + B S_z,
\end{equation}
where $d^\dag_\s$ creates a spin-$\s$ electron, $\e_{\rm d}$
denotes the energy of an electron in the dot, and $U$ describes
the Coulomb correlations between two electrons occupying the dot.
The last term corresponds to external magnetic field $B$ applied
along the $z$th direction ($g\mu_B\equiv 1$) and $S_z =
\frac{1}{2}(d_\uparrow^\dag d_\uparrow - d^\dag_\downarrow
d_\downarrow)$. The tunnel Hamiltonian is given by
\begin{equation}
H_{\rm tun} = \sum_{rk\s} t_{r\s} \left( d_\s^\dag c_{rk\s} +
c_{rk\s}^\dag d_\s \right),
\end{equation}
where $t_{r\s}$ describes the spin-dependent hopping matrix
elements between the dot and ferromagnetic leads. The coupling to
magnetic leads can be described by $\Gamma_{r\s} = \pi \rho_r
|t_{r\s}|^2$, where $\rho_r \equiv\rho$ is the density of states
in the lead $r$. We have thus shifted the whole spin-dependence
into the coupling constants and assumed a flat band of width
$2D$,~\cite{martinekPRL03,ChoiPRL04} where $D\equiv 1$ is set as
the energy unit, if not stated otherwise.

By means of a unitary transformation in the left-right
basis,~\cite{Glazman98,Ng98} one can map the problem of tunneling
through quantum dot coupled to the left and right leads into a
problem where the dot is effectively coupled to a single lead with
a new coupling constant, $\Gamma_\s = \Gamma_{\rm L\s}+\Gamma_{\rm
R\s}$. This can be done by introducing the following symmetric
operators, $\alpha_{k\s} = \tilde{t}_{\rm L\s} c_{{\rm L}k\s} +
\tilde{t}_{\rm R\s} c_{{\rm R} k \s}$, with dimensionless
coefficients $\tilde{t}_{r\s}=t_{r\s}/\sqrt{t^2_{\rm L\s} +
t^2_{\rm R\s}}$. Then, the tunneling Hamiltonian can be written as
\begin{equation}
H_{\rm tun} = \sum_{k\s} \sqrt{\frac{\Gamma_\s}{\pi\rho}} \left(
d_\s^\dag \alpha_{k\s} + \alpha_{k\s}^\dag d_\s \right).
\end{equation}
One can see that now the dot is tunnel-coupled to only one
effective electron reservoir, $H_{\rm FM} = \sum_{k\s}\e_{k\s}
\alpha_{k\s}^\dag \alpha_{k\s}$, with new spin-dependent coupling
constant $\Gamma_\s$. The other parts of the system Hamiltonian,
Eq.~(\ref{Eq:H}), are not affected by this transformation. To
parameterize the spin-dependent couplings we also introduce the
spin polarization of ferromagnetic leads, $p =
(\Gamma_\uparrow-\Gamma_\downarrow) /
(\Gamma_\uparrow+\Gamma_\downarrow)$. The couplings can be then
written in a compact form as, $\Gamma_{\uparrow(\downarrow)} =
(1\pm p)\Gamma$, where $\Gamma = (\Gamma_\uparrow +
\Gamma_\downarrow)/2$. Assuming symmetric coupling strength of the
dot to the leads, the resultant coupling in the antiparallel
configuration is the same for the spin-up and spin-down electrons,
$\Gamma_{\uparrow(\downarrow)}^{\rm AP} = \Gamma$. On the other
hand, in the parallel configuration, the couplings are then
different for the two spin directions,
$\Gamma_{\uparrow(\downarrow)}^{\rm P} = (1\pm p)\Gamma$, which
effectively leads to spin splitting of the dot level and, when
this is the case, the Kondo resonance may become suppressed
because of broken spin degeneracy.~\cite{martinekPRL03,ChoiPRL04}

Finally, the exchange Hamiltonian describing the coupling between
the dot and the second (nonmagnetic) reservoir is given by
\begin{equation}
H_{\rm exch} = \frac{J}{2} \sum_{\s\s'}\sum_k
\vec{S} a^\dag_{k\s}\vec{\s}_{\s\s'} a_{k\s'},
\end{equation}
where $\vec{S} = \frac{1}{2} \sum_{\s\s'} d^\dag_\s
\vec{\s}_{\s\s'} d_{\s'}$ is the spin in the dot, $J$ denotes the
exchange coupling constant and $\vec{\s}$ is a vector of Pauli
spin matrices. We note that in addition to the exchange scattering
of electrons [described by Eq.~(5)] there could be also potential
scattering. However, in this work we are mainly interested in the
low energy physics, where the Kondo effect emerges, so the
potential scattering may be neglected, as it does not lead to any
Kondo-type correlations.

\subsection{Method}

To analyze the equilibrium transport properties of the considered
system, we employ the numerical renormalization group method
\cite{WilsonRMP75} -- nonperturbative, very powerful and
essentially exact numerical method to address quantum impurity
problems. \cite{BullaRMP08} The NRG consists in a logarithmic
discretization of the conduction band and mapping of the system
onto a semi-infinite chain with the impurity (quantum dot) sitting
at the end of the chain. By diagonalizing the Hamiltonian at
consecutive sites of the chain and storing the eigenvalues and
eigenvectors of the system, one can calculate the static and
dynamic quantities of the system. In the case of model considered
in this paper, the Hamiltonian is mapped onto two semi-infinite
chains, where the first chain corresponds to ferromagnetic leads
tunnel-coupled to the dot, while the second one to nonmagnetic
reservoir exchange-coupled to the dot. Because such two-channel
calculations are usually very demanding numerically, it is crucial
to exploit as many symmetries of the system's Hamiltonian as
possible. Especially, using the $SU(2)$ symmetry decreases the
size of Hilbert space and thus increases considerably the accuracy
of calculations. In particular, to efficiently perform the
analysis, we have used the flexible density-matrix numerical
renormalization group (DM-NRG) code, which can tackle with
arbitrary number of both Abelian and non-Abelian
symmetries.~\cite{Toth_PRB08,FlexibleDMNRG} In calculations we
have thus used the $U(1)$ symmetry for the $z$th component of the
total spin, the $U(1)$ symmetry for the charge in the first
channel, and the $SU(2)$ symmetry for the charge in the second
channel. Furthermore, in calculations we have taken the
discretization parameter $\Lambda=2$ and kept 3000 states at each
iteration step.

Using the NRG we can calculate the spectral function of the dot,
$A_\s(\omega) = -\frac{1}{\pi}{\rm Im}G^R_{d\s}(\omega)$, where
$G_{d\s}^R(\omega)$ denotes the Fourier transform of the dot
retarded Green's function, $G_{d\s}^R(t)= -i\Theta(t)
\expect{\{d_{\s}(t), d_{\s}^\dag(0)\}}$. On the other hand, the
spectral function can be directly related to the spin-resolved
linear conductance $G_\s$ by the following formula
\begin{equation}
  G_\s = \frac{e^2}{h}\frac{4\Gamma_{\rm L\s}\Gamma_{\rm R\s}}
  {\Gamma_{\rm L\s} + \Gamma_{\rm R\s}}\int
  d\omega \pi A_\s(\omega)\left(-\frac{\partial
  f(\omega)}{\partial\omega}\right),
\end{equation}
where $f(\omega)$ is the Fermi distribution function and the total
conductance is given by, $G=G_{\uparrow}+G_{\downarrow}$. At zero
temperature, the spin dependent conductance for the parallel
configuration is given by, $G_{\uparrow(\downarrow)}^{\rm P} =
\frac{e^2}{h}(1\pm p)\pi\Gamma A_{\uparrow(\downarrow)}^{\rm P}$,
while for the antiparallel configuration one gets,
$G_{\uparrow(\downarrow)}^{\rm AP} = \frac{e^2}{h}(1-
p^2)\pi\Gamma A_{\uparrow(\downarrow)}^{\rm AP} = G^{\rm AP}/2$,
where $A_\s^{\rm P/AP}$ is the zero-temperature spectral function
of the $d$-level operator in respective magnetic configuration,
taken at $\omega =0$.

\section{Numerical results}

In the following we present numerical results on the equilibrium
spectral function and linear conductance, when the  quantum dot is
in the Kondo regime. We will distinguish between two different
situations; symmetric ($\e_{\rm d}=-U/2$) and asymmetric ($\e_{\rm
d}\neq-U/2$) Anderson models. The origin of such a distinction
stems from the way in which ferromagnetic leads act on the quantum
dot. More specifically, in the asymmetric Anderson model
ferromagnetism of the leads gives rise to a spin splitting of the
Kondo resonance in the parallel configuration, while in the
symmetric model no such a splitting appears (assuming that the dot
is coupled with the same strength to the left and right
leads).~\cite{martinekPRL03,ChoiPRL04,martinek_PRB05} In other
words, an effective exchange field, due to coupling to magnetic
leads, acts on the dot in the former case, while such a field
vanishes in the latter case. The effective field is directly
related to the difference in the coupling strengths of the dot and
ferromagnetic leads for the two spin orientations. Since the
coupling in the spin-up channel is larger than that in the
spin-down one, energy of the spin-up (spin-down) electron in the
dot decreases (increases) by $\Delta\e_{\rm d}/2$. Consequently,
the spin-dependent coupling acts as an effective magnetic field,
leading to spin-splitting $\Delta\e_{\rm d}$ of the dot
level.~\cite{martinekPRL03,ChoiPRL04,martinek_PRB05}

\subsection{Symmetric Anderson model}

\begin{figure}[t]
  \includegraphics[width=0.85\columnwidth]{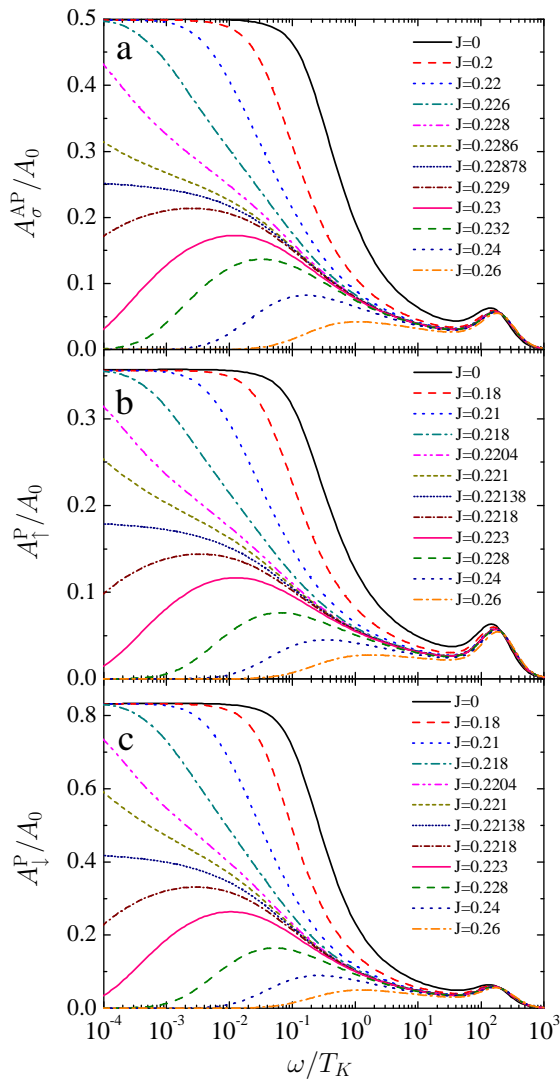}
  \caption{\label{Fig:2}
  (color online) The spectral function of the $d$-level operator
  in the antiparallel (a) and parallel (b,c) magnetic
  configurations for the symmetric Anderson model
  and for different values of the exchange coupling $J$.
  The parameters are:
  $\varepsilon_{\rm d} = -0.05$, $U=0.1$, $\Gamma=0.0077$, $p=0.4$, and $T=0$.
  The Kondo temperature is defined as a half-width
  of the spectral function for $J=0$ and $p=0$, $T_K=2.5\times 10^{-4}$, while
  $A_0 = \sum_\sigma A_\sigma (\omega =0)$ for $J=0$ and $p=0$.
  All the parameters are given in the units of $D\equiv 1$.}
\end{figure}

For the symmetric Anderson model we assume the following
parameters (in the units of $D$), $\varepsilon_{\rm d} = -0.05$
and $U=0.1$. The zero-temperature spin-dependent spectral function
$A_\sigma$, normalized to $A_0$, with $A_0 = \sum_\sigma A_\sigma
(\omega =0)$ taken for $J=0$ and $p=0$, is shown in
Fig.~\ref{Fig:2} for both antiparallel (a) and parallel (b,c)
magnetic configurations (note the logarithmic energy scale), and
for indicated values of the exchange coupling parameter $J$. The
spectral function is plotted as a function of $\omega/T_K$, where
$T_K$ is the Kondo temperature defined as a half-width of the
$d$-level spectral function for $J=0$ and $p=0$, $T_K=2.5\times
10^{-4}$. It can be seen that in the antiparallel configuration
the spectral function is independent of the spin orientation
[Fig.~\ref{Fig:2}(a)], $A_\uparrow^{\rm AP} = A_\downarrow^{\rm
AP}$, while it depends on electron spin in the parallel magnetic
configuration, $A_\uparrow^{\rm P} \neq A_\downarrow^{\rm P}$, see
Fig.~\ref{Fig:2}(b) and (c). Note, that for symmetric Anderson
model the spectral function is symmetric with respect to
$\omega=0$, therefore here it is shown only for positive energies,
i.e. for energies above the Fermi level. Moreover, note also that
the spectral functions are normalized to that for the
corresponding paramagnetic limit ($p=0$ and $J=0$), so $A_\uparrow
(\omega=0) +A_\downarrow (\omega=0) \neq A_0$ in the parallel
configuration.

Let us consider first the situation with vanishing exchange
coupling of the dot to the nonmagnetic reservoir, $J=0$. For
$\omega < T_K$, a Kondo peak develops in the dot spectral function
due to screening of the dot's spin by conduction electrons of
ferromagnetic leads, which leads to the formation of a non-local
spin singlet. The height of the Kondo peak is independent of spin
in the antiparallel configuration and depends on spin in the
parallel one. Apart from this, a Hubbard peak corresponding to
$\varepsilon_{\rm d} + U$ is visible in the spectral function
shown in Fig.~\ref{Fig:2}. This behavior of the Kondo phenomenon
in the presence of ferromagnetic leads is in agreement with that
found by other methods, for instance by the equation of motion for
the Green functions~\cite{SwirkowiczPRB06} and also by the
real-time diagrammatic technique.~\cite{UtsumiPRB05}

\begin{figure}[t]
  \includegraphics[width=0.85\columnwidth]{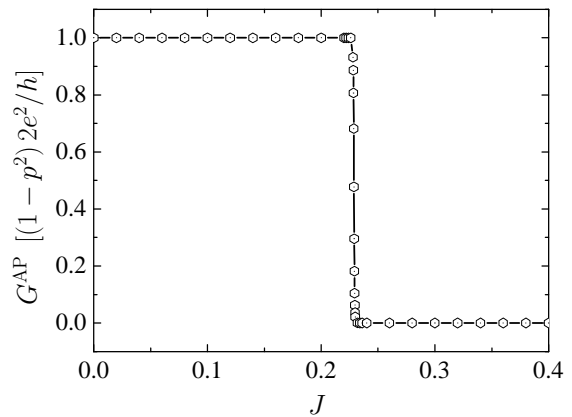}
  \caption{\label{Fig:3}
  The linear conductance as a function of exchange
  coupling constant $J$ for the antiparallel magnetic configuration.
  The conductance was determined from the spectral function shown in
  Fig.~\ref{Fig:2}(a).
  The parameters are the same as in Fig.~\ref{Fig:2} and $J$
  is in units of $D=1$.}
\end{figure}

The situation changes when the electron in the dot is additionally
exchange coupled to the nonmagnetic reservoir. When the coupling
is antiferromagnetic and the coupling parameter $J$ increases, the
width of the Kondo peak becomes gradually narrower and narrower.
The hight of the peak, however, remains unchanged, as can be
clearly seen in Fig.~\ref{Fig:2} for some small values of the
exchange coupling constant. In order to see this behavior also for
larger $J$, but still smaller than a critical value, $J=J_{\rm
c}^{\rm P(AP)}$, one should plot the spectral function for lower
energies. For $J<J_{\rm c}^{\rm P(AP)}$, the system is in the spin
singlet ground state formed by the quantum dot spin and electrons
in the ferromagnetic leads, which gives rise to the Kondo
resonance in the spectral function. However, when $J>J_{\rm
c}^{\rm P(AP)}$, the coupling to nonmagnetic reservoir becomes
larger than the coupling to ferromagnetic leads and the dot's spin
becomes screened by electrons of the nonmagnetic reservoir. Now
the Kondo peak in the spectral function disappears for both
magnetic configurations of the system, see Fig.~\ref{Fig:2}. When
the two couplings are equal, i.e. for $J=J_{\rm c}^{\rm P(AP)}$,
the system is in an exotic state where the two channels try to
screen the dot's spin. The spectral function at $\omega=0$ is then
equal to a half of its value corresponding to $J=0$, $\frac{1}{2}
A^{\rm P/AP}_\sigma |_{J=0}$, for both magnetic configurations,
see Fig.~\ref{Fig:2}. This behavior reveals a quantum phase
transition with increasing strength of the exchange coupling. The
origin of the phase transition follows from the interplay of the
tunnel coupling to the ferromagnetic electrodes and exchange
coupling to the nonmagnetic reservoir. More specifically, the
quantum phase transition occurs at the boundary between two
different singlet ground states, involving the dot's spin and
conduction electrons of the leads or side-coupled reservoir. The
behavior of transport characteristics around this critical point
in the case of nonmagnetic system was discussed in
Ref.~[\onlinecite{PustilnikPRB04}]. It was shown that the
zero-temperature conductance depends step-like on the difference
$\Delta$ between the tunnel and exchange couplings, and becomes
equal to a half of its maximum value at the critical point, i.e.
when $\Delta=0$. The discontinuity of the linear conductance with
respect to $\Delta$ reflects the quantum phase transition in the
parameter space of tunnel coupling $t$ and exchange coupling $J$.
Since the conductance is determined by the corresponding spectral
functions at $\omega=0$, quantum critical behavior is also
reflected in the $J$-dependence of the spectral function. We also
note that at finite temperature the transition is smeared as
$\sqrt{T/T_K}$ and turns rather into a crossover.
\cite{PustilnikPRB04}

\begin{figure}[t]
  \includegraphics[width=0.85\columnwidth]{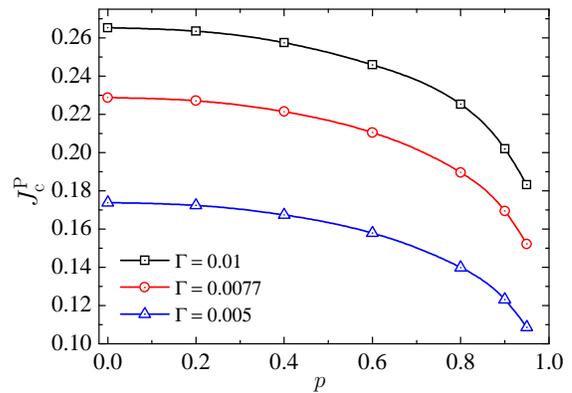}
  \caption{\label{Fig:4}
  (color online) The dependence of the critical exchange
  coupling $J_{\rm c}^{\rm P}$ (in units of $D=1$)
  on the spin polarization of the leads $p$
  in the case of symmetric Anderson model and parallel magnetic
  configuration for three different values
  of the tunnel coupling $\Gamma$, as indicated in the figure.
  The other parameters are the same as in Fig.~\ref{Fig:2}.}
\end{figure}

Using the Schrieffer-Wolff transformation,~\cite{Schrieffer-Wolff}
one could try to estimate the critical value of $J$. For the
symmetric Anderson model and for antiparallel configuration one
gets,~\cite{hewson_book} $J^{\rm S-W}_{\rm c} = \Gamma
(\pi\rho)^{-1} U |\e_{\rm d}|^{-1}(\e_{\rm d}+U)^{-1} \approx
0.196$. From the numerical data, however, one finds $J_{\rm
c}^{\rm AP}\approx 0.22878$ for the antiparallel and $J_{\rm
c}^{\rm P}\approx 0.22138$ for the parallel configurations, see
Fig.~\ref{Fig:2}. The difference between the value obtained using
the Schrieffer-Wolff transformation and the numerical value may
result for example from the fact that the transformation is based
on perturbation expansion, and takes into account only the
second-order tunneling processes.

The quantum critical behavior can be also seen in the dependence
of the linear conductance on the coupling constant $J$, which is
shown in Fig.~\ref{Fig:3} for the antiparallel magnetic
configuration. For $J<J_{\rm c}^{\rm AP}$, the conductance is
$G^{\rm AP} = (1-p^2)2e^2/h$ and drops to zero when $J>J_{\rm
c}^{\rm AP}$. On the other hand, at the quantum critical point
$J=J_{\rm c}^{\rm AP}$, the linear conductance is equal to half of
its value for $J=0$, i.e. $G^{\rm AP} = (1-p^2)e^2/h$.
Consequently, the dependence of the conductance on the exchange
coupling $J$ can be expressed as $G^{\rm AP} =
\Theta(\Delta)(1-p^2)2e^2/h$, where $\Delta=J_{\rm c}^{\rm AP}-J$
and $\Theta(x)$ is the step function. The dependence of $G$ on $J$
for the parallel configuration is qualitatively similar to that in
the antiparallel configuration, therefore it is not shown here.

In Fig.~\ref{Fig:4} the dependence of the critical exchange
coupling $J_{\rm c}^{\rm P}$ on the spin polarization of the leads
$p$ in the case of symmetric Anderson model and parallel magnetic
configuration is shown for three different values of the tunnel
coupling $\Gamma$. First of all, the critical coupling $J_{\rm
c}^{\rm P}$ decreases with decreasing the coupling strength
$\Gamma$. Moreover, $J_{\rm c}^{\rm P}$ also decreases with
increasing the spin polarization of the leads. For $p\to 1$,
$J_{\rm c}^{\rm P}$ tends to zero, as only spins of one
orientation are coupled to the leads and the Kondo effect becomes
suppressed. This behavior of the critical parameter $J_{\rm
c}^{\rm P}$ is consistent with the dependence of the Kondo
temperature in a quantum dot coupled to ferromagnetic leads on the
coupling strength $\Gamma$ and spin polarization
$p$.~\cite{LopezPRL03,martinekPRL03_2,martinekPRL03,ChoiPRL04,sindel_PRB07}

\subsection{Asymmetric Anderson model}

\begin{figure}[t]
  \includegraphics[width=1\columnwidth]{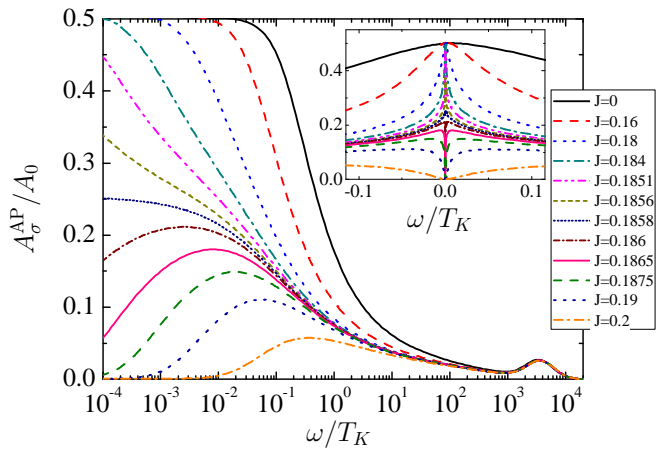}
  \caption{\label{Fig:5}
  (color online) The spectral function of the $d$-level operator
  in the antiparallel magnetic configuration
  for the asymmetric Anderson model, $\varepsilon_{\rm d} = -0.05$, $U=0.2$,
  and for indicated values of the exchange coupling $J$.
  The Kondo temperature for assumed parameters (and for $J=0$ and $p=0$) is $T_K=3.4\times
  10^{-5}$. The other parameters are the same as in Fig.~\ref{Fig:2}.}
\end{figure}

Let us now consider the case of asymmetric Anderson model,
$|\varepsilon_{\rm d}|\ne U/2$. For numerical calculations we
assume $\varepsilon_{\rm d} = -0.05$ and $U=0.2$. The spectral
function in the antiparallel magnetic configuration as a function
of $\omega/T_K$, where $T_K=3.4\times 10^{-5}$, is shown in
Fig.~\ref{Fig:5} for indicated values of the exchange coupling
parameter $J$. The inset shows the behavior of the spectral
function associated with the Kondo peak. The general features of
the spectral function are similar to those of the corresponding
spectral function in the case of symmetric Anderson model
discussed above, see Fig.~\ref{Fig:2}. This is because in the
antiparallel configuration the resultant coupling to ferromagnetic
leads does not depend on spin and the system effectively behaves
as a nonmagnetic one. As before, one observes a quantum phase
transition at $J=J_{\rm c}^{\rm AP}$, where now $J_{\rm c}^{\rm
AP}\approx 0.1858$. The only difference is that for
$|\varepsilon_{\rm d}|\ne U/2$ the spectral function displays an
asymmetric behavior with respect to $\omega=0$, see the inset in
Fig.~\ref{Fig:5}.

\begin{figure}[t]
  \includegraphics[width=0.85\columnwidth]{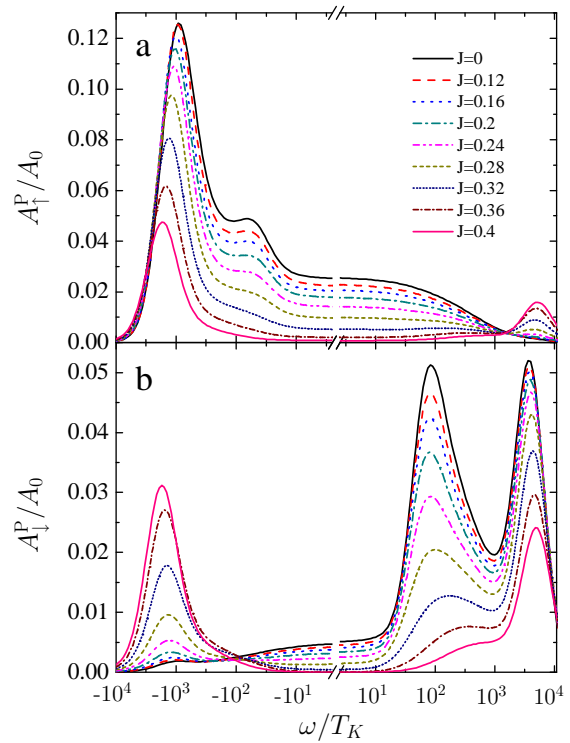}
  \caption{\label{Fig:6}
  (color online) The spectral function of the $d$-level operator
  in the parallel magnetic configuration
  for the asymmetric Anderson model.
  The other parameters are the same as in Fig.~\ref{Fig:5}.}
\end{figure}

The situation, however, changes significantly when the
magnetizations of the leads switch to the parallel configuration.
The corresponding spectral function for spin-$\uparrow$ and
spin-$\downarrow$ is shown in Fig.~\ref{Fig:6}. Note, that now the
spectral function is shown for both positive and negative
energies. As before, let us consider first the case of $J=0$. Due
to an effective exchange field originating from the presence of
ferromagnetic electrodes, the spin degeneracy of the dot level is
lifted. At zero temperature, the magnitude of the splitting due to
exchange field, $\Delta\e_{\rm d}$, can be estimated from the
formula~\cite{martinekPRL03,ChoiPRL04,martinek_PRB05}
\begin{equation}
\Delta\e_{\rm d} = \frac{2p\Gamma}{\pi}\frac{|\e_{\rm
d}|}{|\e_{\rm d}+U|} \,.
\end{equation}
For the assumed parameters one then finds, $\Delta\e_{\rm
d}\approx 2.15\times 10^{-3}$. The exchange field leads generally
to the suppression of the Kondo peak, however the reminiscent of
the Kondo effect are still visible as relatively small peaks in
the $d$-level spectral function. The position of these peaks is
shifted away from the Fermi level -- to positive energies for
spin-$\downarrow$ and to negative energies for spin-$\uparrow$. In
fact, the peaks occur for energies comparable to the magnitude of
the exchange field. As can be seen in Fig.~\ref{Fig:6}, they
develop at $\omega/T_K \approx 10^2$ for spin-$\downarrow$ and at
$\omega/T_K \approx -10^2$ for spin-$\uparrow$ components of the
spectral function. The other peaks in the spectral functions
correspond to the dot level $\e_{\rm d}$ and its Coulomb
counterpart $\e_{\rm d}+U$.

\begin{figure*}
  \includegraphics[height=8.4cm]{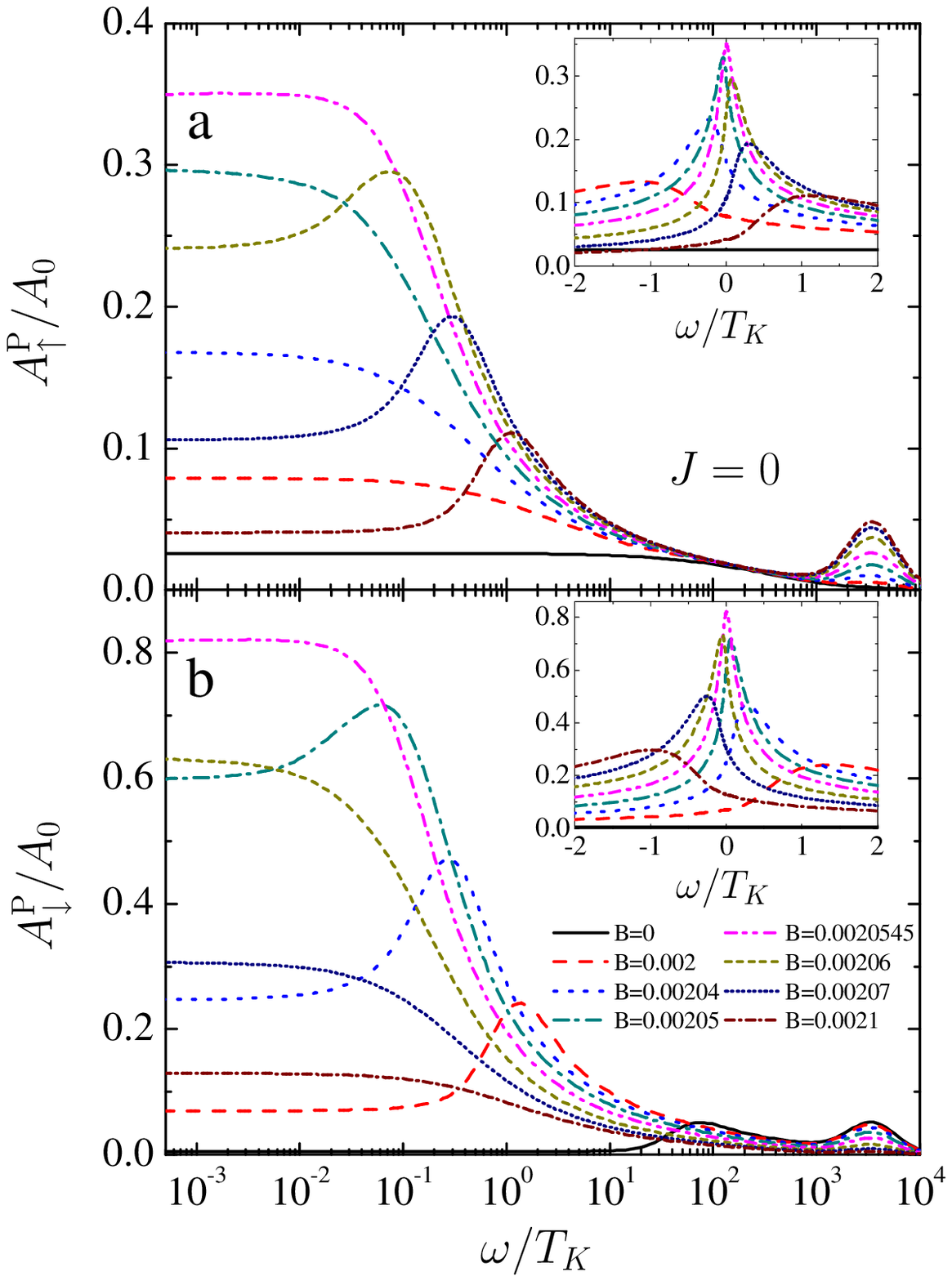}
  \includegraphics[height=8.4cm]{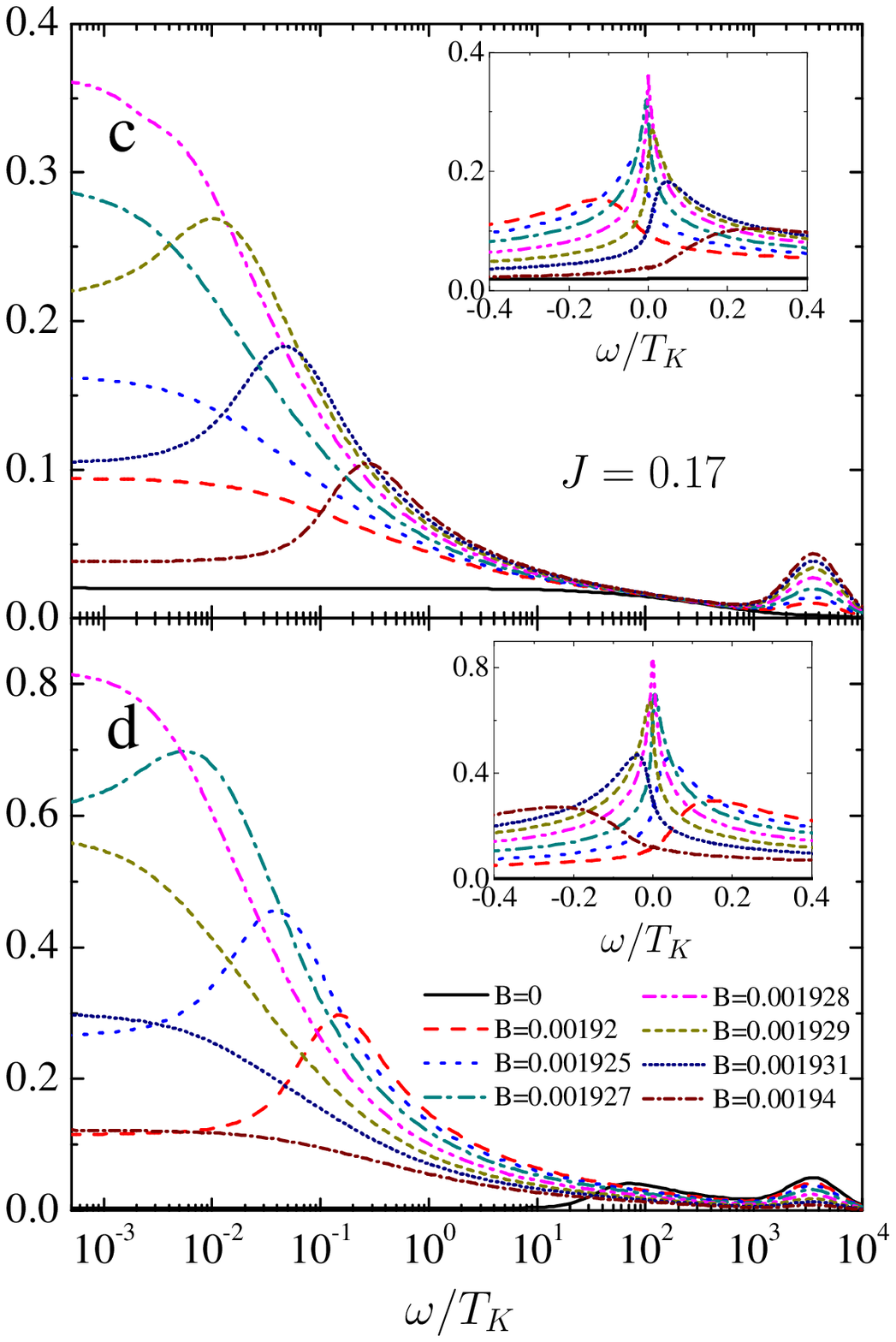}
  \includegraphics[height=8.4cm]{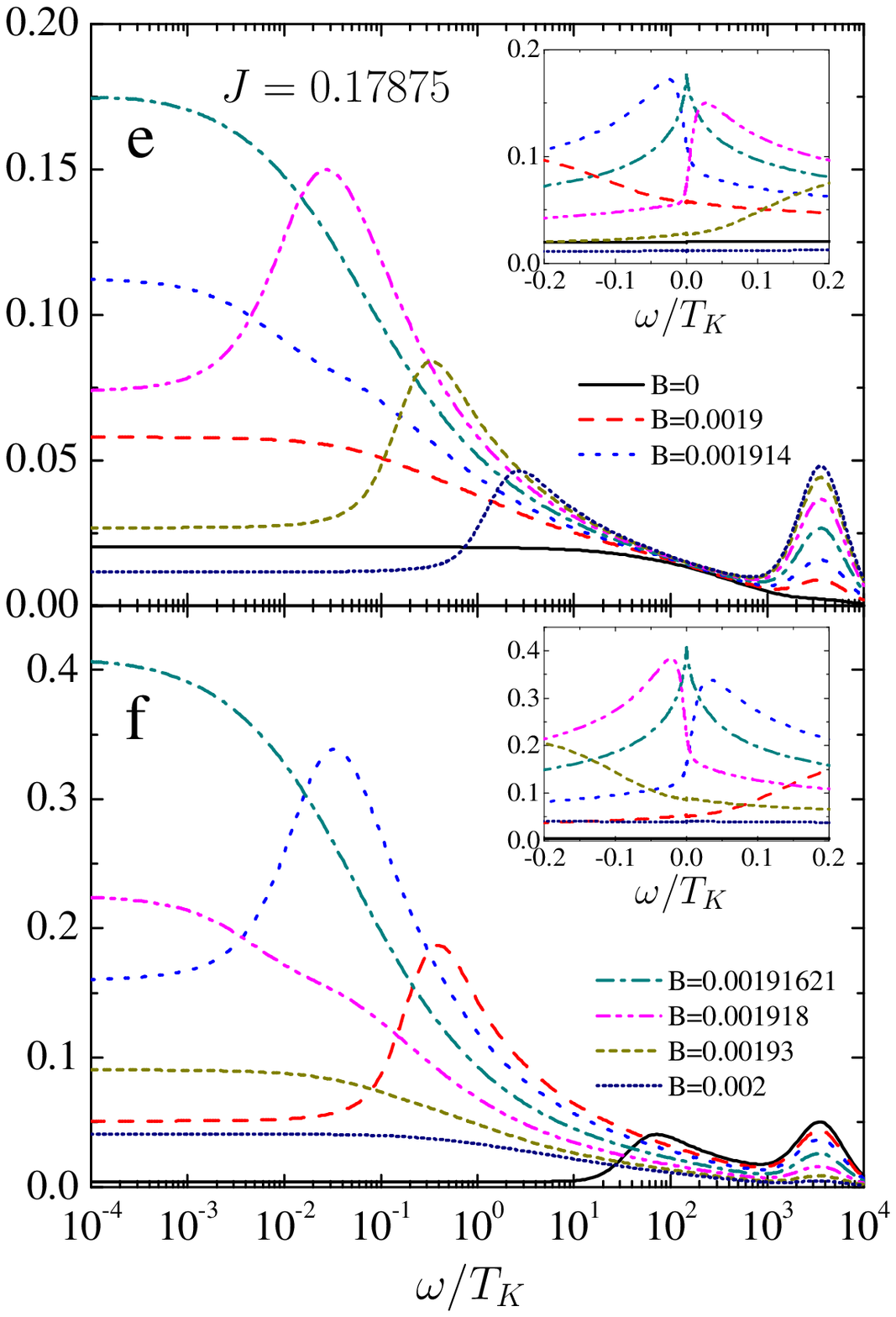}
  \caption{\label{Fig:7}
  (color online) The spectral function of the $d$-level operator
  in the parallel magnetic configuration
  for the asymmetric Anderson model in the presence of external
  magnetic field $B$ applied along the $z$th direction
  for different values of exchange coupling constant
  $J=0$ (a,b), $J=0.17$ (c,d) and $J=0.17875$ (e,f).
  The other parameters are the same as in Fig.~\ref{Fig:5}.}
\end{figure*}

When the coupling parameter $J$ increases, the weak Kondo
resonances in the spectral function gradually disappear for both
spin orientations. The physics behind this disappearance remains
similar to that described above, i.e. screening of the dot's spin
by the nonmagnetic reservoir exchange-coupled to the dot.
Interestingly, there is no quantum phase transition in the case of
parallel magnetic configuration shown in Fig.~\ref{Fig:6}.

Let us now assume that there is an external magnetic field $B$
applied to the dot along the $z$th direction. In the case of
antiparallel configuration, the magnetic field destroys both the
Kondo resonance and quantum phase transition with changing $J$.
However, when the leads are aligned in parallel, the Kondo effect
is already suppressed by the effective exchange field coming from
ferromagnetic electrodes, and one may consider the possibility of
restoring the Kondo peak by applying an external magnetic field
which compensates the effects due to exchange field. In
Fig.~\ref{Fig:7} we show the spectral functions for the parallel
magnetic configuration in the case of an asymmetric Anderson
model, calculated for three different values of the exchange
constant $J$ in the presence of external magnetic field $B$. The
insets display behavior of the spectral function associated with
the Kondo peaks. When $J=0$, see  Fig.~\ref{Fig:7}(a) and (b), the
full Kondo peak at the Fermi level in the spectral density can be
restored for both spin orientations by properly tuned external
magnetic field, which happens for $B=B_{\rm c} = 0.0020545$, where
$B_{\rm c}$ (in the units of D) denotes the compensating field.
This is in agreement with the result obtained
earlier.~\cite{martinekPRL03,ChoiPRL04} Similar behavior also
appears for larger positive $J$, e.g. for $J=0.17$ shown in
Fig.~\ref{Fig:7}(c) and (d). Now, the Kondo resonance becomes
restored when the compensating field is $B_{\rm c} = 0.001928$.
Note that the magnitude of magnetic field necessary for full
restoration of the Kondo effect slightly decreases as $J$
increases. The question which arises now is whether such a
restoration by magnetic field is also possible for larger values
of $J$. By fine-tuning in the parameter space of $J$ and $B$, we
have found that this is the case for $J$ below a certain critical
value $J<\tilde{J}_{\rm c}^{\rm P}= 0.17875$. Here $\tilde{J}_{\rm
c}^{\rm P}$ denotes the critical value of $J$ in the parallel
configuration and in the presence of the compensating magnetic
field. From numerical results (not shown here), follows that for
$J>\tilde{J}_{\rm c}^{\rm P}$, the magnetic field can only
partially restore the Kondo effect, leading to small side peaks in
the spectral function, while the full Kondo peak at $\omega=0$
cannot be restored. One may now expect that for $J=\tilde{J}_{\rm
c}^{\rm P}$, the magnetic field should also restore the quantum
critical state. Indeed, by fine-tuning in the parameter space we
have found that the quantum critical state can be recovered for
$B_{\rm c} = 0.00191621$. This situation is shown explicitly in
Fig.~\ref{Fig:7}(e) and (f). Thus, we have shown that in the
parallel configuration the properly-tuned magnetic field can
restore both the full Kondo effect for $J<\tilde{J}_{\rm c}^{\rm
P}$ as well as the quantum critical state for $J=\tilde{J}_{\rm
c}^{\rm P}$.

\begin{figure}[h]
  \includegraphics[width=0.8\columnwidth]{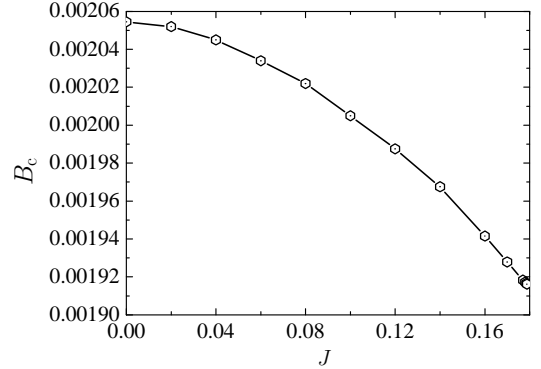}
  \caption{\label{Fig:8}
  The dependence of the compensating magnetic field $B_{\rm c}$
  on the exchange coupling constant $J$ in the case
  of the parallel configuration and asymmetric Anderson model.
  The other parameters are the same as in Fig.~\ref{Fig:5}.
  $J$ and $B_{\rm c}$ are in units of $D=1$.}
\end{figure}

By comparing numerical curves presented in Fig.~\ref{Fig:7}, one
can note that the compensating field $B_{\rm c}$ decreases with
increasing the exchange coupling $J$. This is explicitly shown in
Fig.~\ref{Fig:8}, where we have calculated the dependence of
$B_{\rm c}$ on the exchange coupling $J$. For $J<\tilde{J}_{\rm
c}^{\rm P}$, the Kondo resonance can be fully restored by applying
compensating field $B_{\rm c}$. On the other hand, when
$J>\tilde{J}_{\rm c}^{\rm P}$, the magnetic field cannot
compensate the exchange field,  so the notion of compensating
field becomes meaningless.

\section{Conclusions}

In this paper we have considered spectral and transport properties
of a single-level quantum dot connected to external ferromagnetic
leads and exchange-coupled to a nonmagnetic reservoir. Using the
numerical renormalization group method we have calculated the
zero-temperature $d$-level spectral function and the conductance
through the dot. We have shown that in the antiparallel
configuration, depending on the strength of the exchange
interaction $J$, the Kondo singlet ground state can form, in which
the conduction electrons either in the ferromagnetic leads or in
the nonmagnetic reservoir are involved. In the former case, the
conductance is maximum, whereas in the latter case the conductance
becomes fully suppressed. For a certain critical value of $J$,
$J_{\rm c}^{\rm AP}$, both electron channels try to screen the
dot's spin and the conductance is equal to a half of its maximum
value. The boundary between the two ground states is a quantum
phase transition.

In the parallel magnetic configuration, on the other hand, the
Kondo effect is generally destroyed due to an effective spin
splitting of the dot level caused by the presence of ferromagnetic
leads. However, there are still small side peaks -- reminiscent of
the Kondo effect -- which occur on both sides of the Fermi level
for energies of the order of effective exchange field.
Nevertheless, with increasing the exchange constant $J$, these
peaks become suppressed.

We have also considered the influence of an external magnetic
field on the $d$-level spectral function and shown that in the
parallel configuration the Kondo effect can be restored by
applying appropriately tuned compensating magnetic field for
$J<\tilde{J}_{\rm c}^{\rm P}$, where $\tilde{J}_{\rm c}^{\rm P}$
is the critical value of $J$ in the compensating magnetic field.
If, however, $J>\tilde{J}_{\rm c}^{\rm P}$, the full Kondo effect
cannot be restored by a magnetic field. In addition, we have found
that the quantum critical behavior, which is suppressed in the
parallel configuration, can also be recovered by tuning the
external magnetic field.


\begin{acknowledgments}
This work was supported by funds of the Polish Ministry of Science
and Higher Education as a research project for years 2006-2009.
I.W. also acknowledges support from the Alexander von Humboldt
Foundation, the Foundation for Polish Science and funds of the
Polish Ministry of Science and Higher Education as a research
project for years 2008-2010. Financial support by the Excellence
Cluster "Nanosystems Initiative Munich (NIM)" is gratefully
acknowledged.
\end{acknowledgments}



\begin{thebibliography}{99}

\bibitem{goldhaber-gordon_98}
D. Goldhaber-Gordon, H. Shtrikman, D. Mahalu, D. Abusch-Magder, U.
Meirav, and M. A. Kastner, Nature (London) 391, 156 (1998).

\bibitem{cronenwett_98}
S. Cronenwett, T. H. Oosterkamp, and L. P. Kouwenhoven, Science
281, 182 (1998).

\bibitem{hewson_book}
A. C. Hewson, {\it The Kondo Problem to Heavy Fermions} (Cambridge
University Press, Cambridge, 1993).


\bibitem{LopezPRL03} 
Rosa Lopez and David Sanchez, Phys. Rev. Lett. {\bf 90}, 116602
(2003).

\bibitem{martinekPRL03_2}
J. Martinek, Y. Utsumi, H. Imamura, J. Barna\'s, S. Maekawa, J.
K\"onig, and G. Sch\"on, Phys. Rev. Lett. {\bf 91}, 127203 (2003).

\bibitem{martinekPRL03}
J. Martinek, M. Sindel, L. Borda, J. Barna\'s, J. K\"onig, G. Sch\"on, and J. von Delft,
Phys. Rev. Lett. {\bf 91}, 247202 (2003).

\bibitem{ChoiPRL04}
Mahn-Soo Choi, David Sanchez, and Rosa Lopez, Phys. Rev. Lett.
{\bf 92}, 056601 (2004).

\bibitem{MatsubayashiPRB07}  
Daisuke Matsubayashi and Mikio Eto, Phys. Rev. B {\bf 75}, 165319
(2007).

\bibitem{SimonPRB07} 
P. Simon, P. S. Cornaglia, D. Feinberg, and C. A. Balseiro, Phys.
Rev. B {\bf 75}, 045310 (2007).

\bibitem{martinek_PRB05}
J. Martinek, M. Sindel, L. Borda, J. Barna\'s, R. Bulla, J.
K\"onig, G. Sch\"on, S. Maekawa, J. von Delft, Phys. Rev. B {\bf
72}, 121302(R) (2005).

\bibitem{sindel_PRB07}
M. Sindel, L. Borda, J. Martinek, R. Bulla, J. K\"onig, G.
Sch\"on, S. Maekawa, and J. von Delft, Phys. Rev. B {\bf 76},
045321 (2007).

\bibitem{pasupathy_04}
A. N. Pasupathy, R. C. Bialczak, J. Martinek, J. E. Grose, L. A.
K. Donev, P. L. McEuen, and D. C. Ralph, Science 306, 86 (2004).

\bibitem{heersche_PRL06}
H. B. Heersche, Z. de Groot, J. A. Folk, L. P. Kouwenhoven, H. S.
van der Zant, A. A. Houck, J. Labaziewicz, and I. L. Chuang, Phys.
Rev. Lett. 96, 017205 (2006).

\bibitem{hamayaAPL07}
K. Hamaya, M. Kitabatake, K. Shibata, M. Jung, M. Kawamura, K.
Hirakawa, T. Machida, T. Taniyama, S. Ishida and Y. Arakawa, Appl.
Phys. Lett. {\bf 91}, 232105 (2007).

\bibitem{hamaya_PRB08}
K. Hamaya, M. Kitabatake, K. Shibata, M. Jung, M. Kawamura, S.
Ishida, T. Taniyama, K. Hirakawa, Y. Arakawa, and T. Machida,
Phys. Rev. B 77, 081302(R) (2008).

\bibitem{hauptmann_NatPhys08}
J. Hauptmann, J. Paaske, P. Lindelof, Nature Phys. 4, 373 (2008).

\bibitem{parkinNL08}
H. Yang, S.-H. Yang, S. S. P. Parkin, Nano Lett. {\bf 8}, 340
(2008).

\bibitem{OregPRL03}
Y. Oreg and D. Goldhaber-Gordon, Phys. Rev. Lett. {\bf 90}, 136602
(2003).

\bibitem{Nozieres_JP80}
P. Nozieres and A. Blandin, J. Phys. {\bf 41}, 193 (1980).

\bibitem{Zawadowski_PRL80}
A. Zawadowski, Pys. Rev. Lett. {\bf 45}, 211 (1980).

\bibitem{AffleckPRB93}
I. Affleck, A. W. W. Ludwig, Phys. Rev. B {\bf 48}, 7297 (1993).

\bibitem{RalphPRL94}
D. C. Ralph, A. W. W. Ludwig, J. von Delft, R. A. Buhrman, Phys.
Rev. Lett. {\bf 72}, 1064 (1994).

\bibitem{HettlerPRL94}
M. H. Hettler, J. Kroha, S. Hershfield, Phys. Rev. Lett. {\bf 73},
1967 (1994).

\bibitem{AndreiPRL95}
N. Andrei and A. Jerez, Phys. Rev. Lett. {\bf 74}, 4507 (1995).

\bibitem{LebanonPRB03}
E. Lebanon, A. Schiller, F. B. Anders, Phys. Rev. B {\bf 68},
155301 (2003).

\bibitem{FlorensPRL04}
S. Florens, A. Rosch, Phys. Rev. Lett. {\bf 92}, 216601 (2004).

\bibitem{PotokNature07}
R. M. Potok, I. G. Rau, Hadas Shtrikman, Yuval Oreg, and D.
Goldhaber-Gordon, Nature {\bf 446}, 167 (2007).

\bibitem{PustilnikPRB04}
M. Pustilnik, L. Borda, L. I. Glazman, and J. von Delft, Phys.
Rev. B {\bf 69}, 115316 (2004).

\bibitem{tothPRB07}
A. I. T\'{o}th, L. Borda, J. von Delft, and G. Z\'{a}rand, Phys.
Rev. B {\bf 76}, 155318 (2007).

\bibitem{LiuJPCM08}
Y. S. Liu X. F. Yang, X. H. Fan and Y. J. Xia, J. Phys.: Condens.
Matter {\bf 20}, 135226 (2008).

\bibitem{Glazman98}
L. I. Glazman and M. E. Raikh, JETP Lett. {\bf 47}, 452 (1988).

\bibitem{Ng98}
T. K. Ng and P. A. Lee, Phys. Rev. Lett. {\bf 61}, 1768 (1988).

\bibitem{WilsonRMP75}
K. G. Wilson, Rev. Mod. Phys. {\bf 47}, 773 (1975).

\bibitem{BullaRMP08}
R. Bulla, T. A. Costi, and T. Pruschke Rev. Mod. Phys. {\bf 80},
395 (2008).

\bibitem{Toth_PRB08}
A. I. T\'{o}th, C. P. Moca, O. Legeza, and G. Zar\'{a}nd, Phys.
Rev. B {\bf 78}, 245109 (2008).

\bibitem{FlexibleDMNRG}
O. Legeza, C. P. Moca, A. I. T\'{o}th, I. Weymann, G. Zar\'{a}nd,
arXiv:0809.3143 (2008) (unpublished).

\bibitem{SwirkowiczPRB06}
R. \'Swirkowicz, M. Wilczy\'nski, M. Wawrzyniak, and J. Barna\'s,
Phys. Rev. B {\bf 73}, 193312 (2006).

\bibitem{UtsumiPRB05}
Y. Utsumi, J. Martinek, G. Sch\"on, H. Imamura, and S. Maekawa,
Phys. Rev. B {\bf 71}, 245116 (2005).

\bibitem{Schrieffer-Wolff}
J. R. Schrieffer and P. A. Wolff, Phys. Rev. {\bf 149}, 491
(1966).

\end{thebibliography}
\end{document}